\documentclass[sn-mathphys-num]{sn-jnl}

\usepackage{graphicx}%
\usepackage{multirow}%
\usepackage{amsmath,amssymb,amsfonts}%
\usepackage{amsthm}%
\usepackage{mathrsfs}%
\usepackage[title]{appendix}%
\usepackage{xcolor}%
\usepackage{textcomp}%
\usepackage{manyfoot}%
\usepackage{booktabs}%
\usepackage{algorithm}%
\usepackage{algorithmicx}%
\usepackage{algpseudocode}%
\usepackage{listings}%

\usepackage{tikz}
\usepackage{float}
\usetikzlibrary{arrows}
\usepackage{units}
\usepackage{setspace}
\usepackage{tabularx}
\usepackage{booktabs}
\usepackage{subfig}
\usepackage{todonotes}
\usepackage{nicefrac}
\usepackage{color}
\usepackage{natbib}
\usepackage{upgreek}

\theoremstyle{thmstylethree}%
\definecolor{dgreen}{RGB}{26,102,46}
\definecolor{orange}{RGB}{255,140,0}
\definecolor{violet}{RGB}{138,43,226}

\begin{document}

\title[Lattice structures: energy absorption at different volume fractions under compressive loading] {Energy absorption of sustainable lattice structures under static compression}


\author*[1]{\fnm{S\"oren} \sur{Bieler}}\email{soeren.bieler@uni-siegen.de}

\author[1]{\fnm{Kerstin} \sur{Weinberg}}\email{kerstin.weinberg@uni-siegen.de}

\affil*[1]{\orgdiv{Department of Mechanical Engineering, Festk\"orpermechanik}, \orgname{Universit\"at Siegen}, \orgaddress{\street{Paul-Bonatz-Stra{\ss}e 9-11}, \city{Siegen}, \postcode{57076}, \state{NRW}, \country{Germany}}}
%
%


\abstract{Lattice-like cellular materials, with their unique combination of lightweight, high strength, and good deformability, are promising for engineering applications.

This paper investigates the energy-absorbing properties of four truss-lattice structures with two defined volume fractions of material in static compression experiments. The mass-specific energy absorption is derived. The specimens are manufactured by SLA printing of viscoelastic polymeric material. Sustainability implies that the lattice  structures can withstand multiple loads and return to their original state after some recovery.

Additionally, we present finite element simulations of our experiments and show that these calculations are, in principle, able to predict the different responses of the lattices. Like in the experiments, the truncated octahedron-lattice structure proved to be the most effective for energy absorption under strong  compression.
}




\maketitle

\section{Introduction}\label{sec_introduction}

Cellular materials can be found in many areas of nature and engineering.
They are used as a lightweight material,  damping support, or simply as a soft base. 
Many of these materials are able to undergo large deformations under mechanical load and yet return to their original state when the load is removed. This makes cellular materials particularly suitable for non-destructive, reversible and, therefore, sustainable energy absorption.

Natural cellular materials are, for example, tree bark, sponges and cork. Here, the cell structure is irregular and non-uniform and, therefore, difficult to reproduce for engineering applications. For the latter, regular arrangements such as honeycomb-filled laminates and hierarchical honeycombs \cite{chen20183d} have become established, but lattice-like structures are also employed. 

\begin{figure*} 
\centering
\resizebox{1\textwidth}{!}{
\begin{tikzpicture}
\draw (0,0) node[anchor=south west]{
\includegraphics[width=1\textwidth]{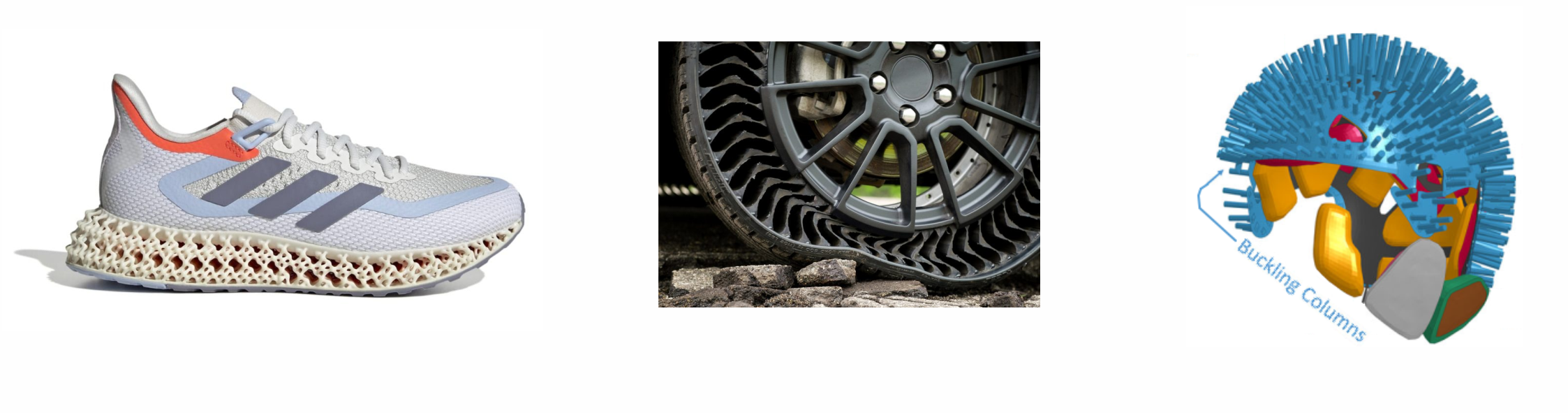}};
\end{tikzpicture}}
\caption{Lattice structures in different applications: shoe sole \textsc{adidas 4DFWD} (left) \cite{adidas2021website}, car tire developed by \textsc{Michelin} (center) \cite{michelin2019wheel} and the internal structure of a modern NFL helmet (right) \cite{Giudice2020}}
\label{fig:einsatz_lattices}
\end{figure*}

Some recent applications are shown in Fig.~\ref{fig:einsatz_lattices}, namely 
an adapted  lattice based  sole of an \textsc{ADIDAS} running shoe \cite{adidas2021website}, 
a new lightweight car tire \cite{michelin2019wheel},
and the interior of specially designed football helmets \cite{Giudice2020}, which are supposed to withstand multiple impacts without losing their protective effect.

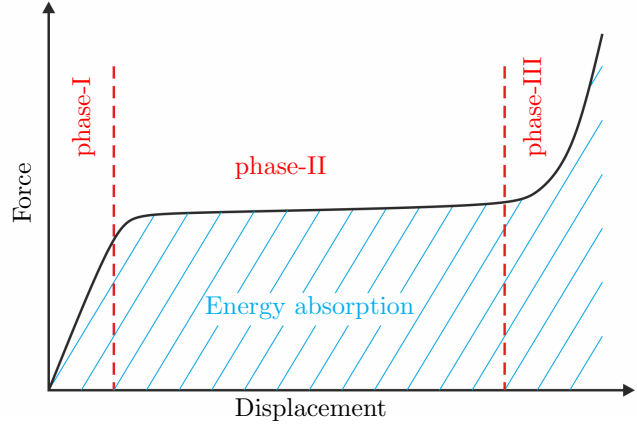
\begin{figure*} 
\centering
\resizebox{1\textwidth}{!}{
\begin{tikzpicture}
\draw (0,0) node[anchor=south west]{
\includegraphics[width=1\textwidth]{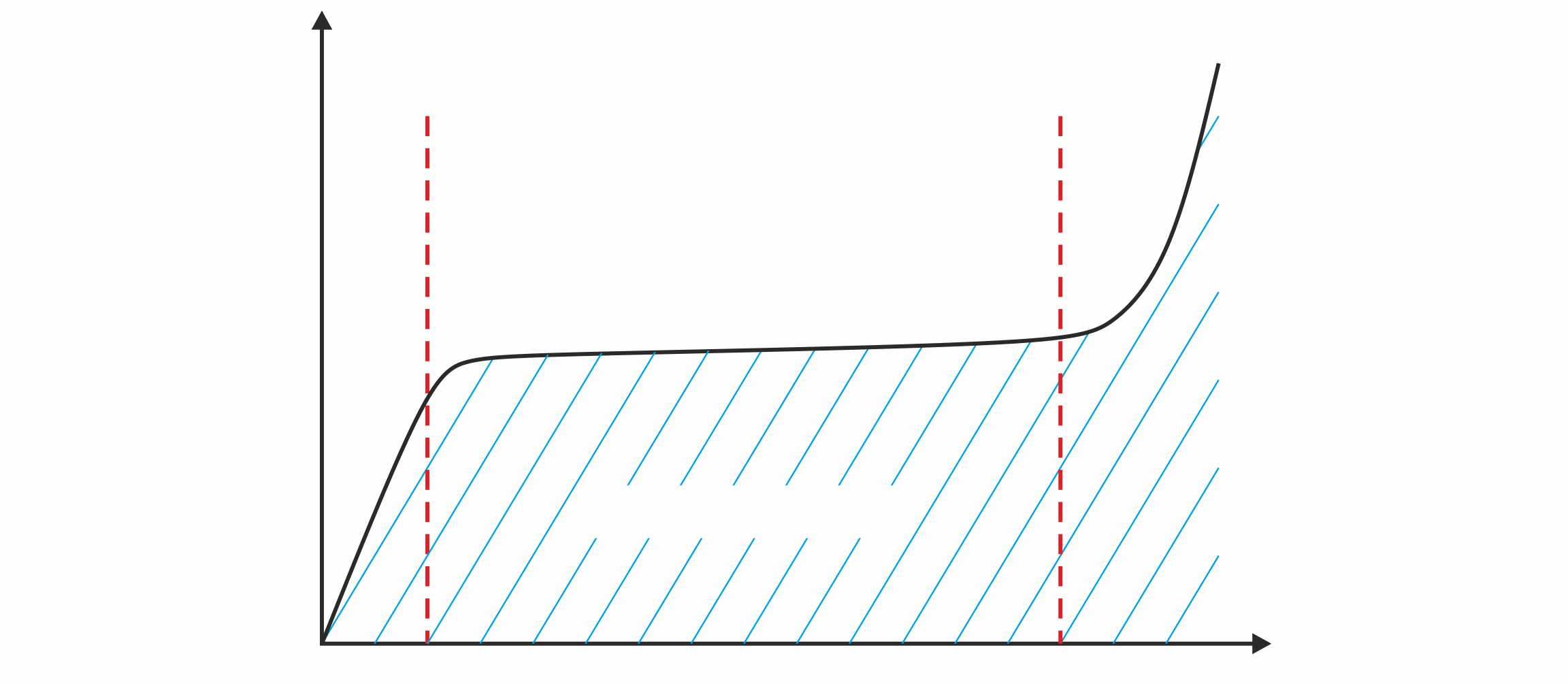}};
\draw(2.2,2.6) node[anchor=south west]{\begin{sideways}Force\end{sideways}};
\draw(5.2,-0.1) node[anchor=south west]{Displacement};
\draw(4.8,1.3) node[anchor=south west]{\textcolor{cyan}{Energy absorption}};
\draw(3,3.5) node[anchor=south west]{\begin{sideways}\textcolor{red}{phase-I}\end{sideways}};
\draw(5.2,3.2) node[anchor=south west]{\textcolor{red}{phase-II}};
\draw(9.1,3.5) node[anchor=south west]{\begin{sideways}\textcolor{red}{phase-III}\end{sideways}};
\end{tikzpicture}}
\caption{Typical force-displacement diagram for cellular structures under compression}
\label{fig:typical_force_displacement}
\end{figure*}

Cellular  materials, in particular foam-like materials,  are described in detail in \cite{ashby1997cellular}. When compressed, they behave as shown schematically in Fig.~\ref{fig:typical_force_displacement}. Initially, the individual cells deform linear-elastically and are relatively stiff (phase-I). When the individual struts fail and begin to buckle, the load-bearing capacity does not increase any further. A plateau is formed (phase-II). At very high compression, when all cells are closed, the resistance to deformation is determined by the ground material and is correspondingly high (phase-III). 

With additive manufacturing, it has become possible to create structures of almost any complexity that would be very difficult or costly to produce using conventional manufacturing processes, cf. \cite{gibson2021design}. This also raises the question of whether much tougher cellular structures than  foam can be constructed, inspired, e.g. by the crystal lattice properties of hexagonally packed or body-centered cubic metals. Numerous such lattice structures have meanwhile been designed and  mechanically tested; for a state of research and a classification of lattices into subcategories, we refer to the review \cite{yin2023review}.

With this in mind, we investigate  here the reversible and thus sustainable energy absorption of four lattice structures.  The acrylic-like polymer material used behaves viscoelastically, similar to polymer foams. Consequently, the lattice structures can withstand multiple loads and return to their original state after a certain recovery time. The structures investigated are beam-based and truss-like, i.e. they are  made up of beam elements. In order to produce a lattice structure from unit cells, they are arranged periodically to form a three-dimensional structure. The investigated lattices  have a defined mass per unit cell. Thus, we compare lattice types with different node connectivity because the beam arrangement and the connectivity influence the mechanical response of the  structure \cite{tancogne2018elastically,deshpande2001foam}.

As an experimental setup, we choose a compression test with up to 50\,\% (quasi)static compression. In the recorded force-displacement diagram, the different answers of the lattice structures become apparent, and we determine the specific energy absorption from the measured data. A higher energy absorption is likely related to a steeper plateau phase in the diagram, i.e., if an increase in force is necessary to compress further.

Our investigations are similar to those in \cite{ling2019mechanical}, where the energy absorption properties under static load of SLA printed  \textsc{Octet} structures in different volume fractions are compared. However, a comparison with other structures, and thus different node connectivities at the same volume fraction $f_\text{V}$, is not given. Other works, such as those by \cite{tancogne2016additively} and \cite{zhong2019mechanical}, investigate energy absorption of metallic lattice structures produced by selective laser melting (SLM). In  \cite{yan2019mechanical}, the authors conduct static compression tests of metallic lattice structures to determine the modulus of elasticity. In the investigations by \cite{xiao2023improving}, body-centered-cubic structures with different beam cross sections manufactured using the SLM process are compressed. We also investigated the properties of micro-scale copper lattice structures under different strain rates \cite{kang2023green} and their energy absorption under dynamic loading \cite{bieler2021investigation}. Due to the metal material, however, the lattices are plastically deformed and thus permanently damaged. For frequent and sustainable use, however, they should be able to withstand multiple loads.

Complementary, to assess the extent to which numerical simulations can replace the relatively complex experiments, we also carry out finite element calculations (FEA). For this purpose, the material properties of the polymer are first determined experimentally. Then, the specimens are meshed in their original size, compressed with up to 50\,\%, and  the force-displacement diagrams are compared with the experimental data.

Our paper is organized as follows: Next, in Section~\ref{sec:lattice_structure_design},  we introduce the investigated lattice structures, discuss their geometry, material, and the manufacturing process by 3D printing.
In Section~\ref{sec:chapter_Experiments} we describe our compression experiments and
present our results on the energy absorption properties. 
In the following Section~\ref{chapter_Simulation} we present the FEA of our experiment.
In a short conclusion in Section~\ref{sec:conclusion} we summarize our findings.

\section{Lattice structure design and manufacturing} \label{sec:lattice_structure_design}

The lattice structures investigated here are truss-like structures made of a polymeric mixture based on acrylates and acrylic acid. The structures we tested are so-called 1st-order structures composed of individual truss elements. The authors \cite{meza2015resilient} have also researched higher-order structures. Here, the truss elements themselves consist of structures.


\subsection{Lattices and structure design}
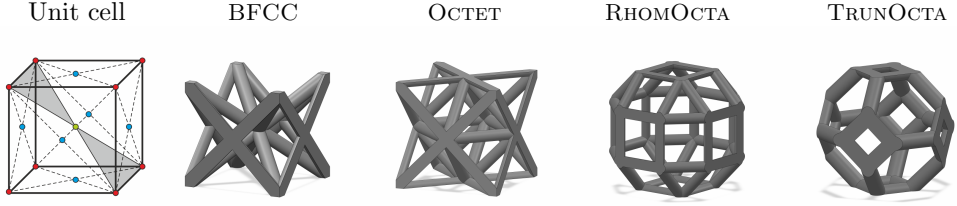
\begin{figure*} 
\centering
\resizebox{1\textwidth}{!}{
\begin{tikzpicture}
\draw (0,0) node[anchor=south west]{
\includegraphics[width=1\textwidth]{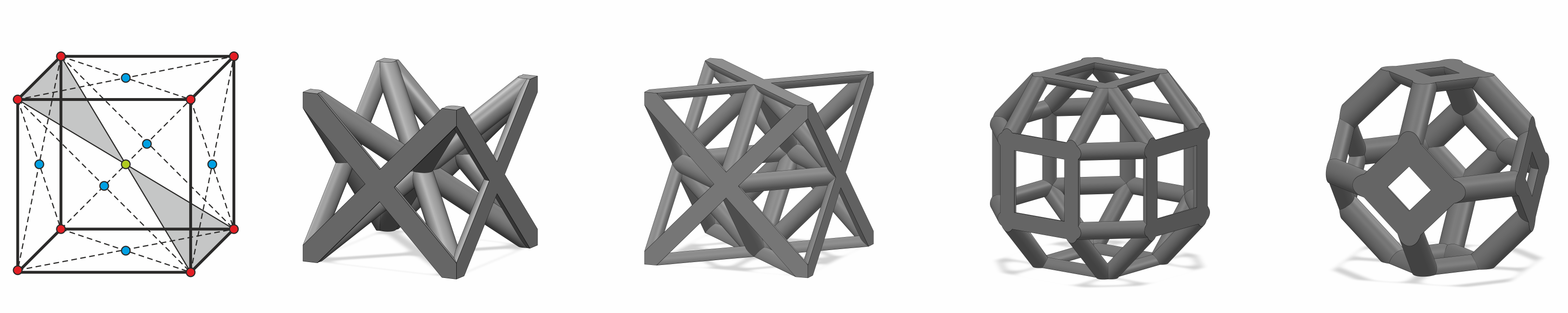}};
\draw(0.4,2.7) node[anchor=south west]{Unit cell};
\draw(3.1,2.7) node[anchor=south west]{\small{\textsc{BFCC}}};
\draw(5.8,2.7) node[anchor=south west]{\small{\textsc{Octet}}};
\draw(8.2,2.7) node[anchor=south west]{\small{\textsc{RhomOcta}}};
\draw(11.2,2.7) node[anchor=south west]{\small{\textsc{TrunOcta}}};
\end{tikzpicture}}
\caption{Unit cells of the investigated lattice structures}
\label{fig:unit_cells}
\end{figure*}

For specifically created lattice  structures, the struts' arrangements and the number of struts per node are subject to design.  Also, the nodes can be arranged at different positions of the unit cell. In this investigation, out of an infinite number of possible lattice designs, two   crystal-like lattices and two convex-truss lattices are chosen for mechanical testing:
\begin{itemize}
  \item a common octet truss lattice composed of tetrahedra and octahedra: \textsc{Octet}
  \item a face-centered cubic lattice with an additional body-centered node: \textsc{BFCC}
  \item a  convex small rhombicuboctahedron truss: \textsc{RhomOcta} 
  \item a  convex polyhedra truss arising from a octahedron by cutting of the vertices (truncated octahedron): \textsc{TrunOcta}
\end{itemize} 
The corresponding unit cells are displayed in Fig.\ref{fig:unit_cells}. Table \ref{tab:struts_nodes} lists the number and positions of nodes and struts of each of the investigated lattice types.

The lattice unit cells differ in the number and positions of nodes and the horizontal and/or declined  arrangement of the struts. In particular,  we expect the convex or 'open' lattices \textsc{RhomOcta} and \textsc{TrunOcta}, with no inner node, to behave differently than the more 'mesh-like'  \textsc{Octet} and \textsc{BFCC} lattices. 

\begin{table*}[htb]
\caption{Number of struts and nodes at different positions of the lattice unit cells}
\label{tab:struts_nodes} 
\begin{tabularx}{\textwidth}{lXXXXr} \toprule
\noalign{\smallskip}
 & struts &  & nodes & \\
 \noalign{\smallskip}
 &  & corner & center & face & $M$\\
\noalign{\smallskip}
\hline
\noalign{\smallskip}
\textsc{BFCC} & 24 & 8 & 1 & 4 & -9\\
\noalign{\smallskip}
\textsc{Octet} & 36 & 8 & - & 6 & 0\\
\noalign{\smallskip}
\textsc{RhomOcta} & 48 & - & - & 24 & -18\\
\noalign{\smallskip}
\textsc{TrunOcta} & 36 & - & - & 24 & -30\\
\noalign{\smallskip}
\hline
\end{tabularx}
\end{table*}

The topological characteristics can in part be described by 
the Maxwell number of the lattices, cf. \cite{deshpande2001effective,leary2018inconel}. The Maxwell number $M$, which classifies the dominating loading regime, is calculated from the $n$ nodes and the $s$ struts which are connected on the nodes, 
\begin{align}
\label{eq:maxwell}
    M = s - 3\,n + 6  \,.
\end{align}
%
%
For $M \geq 0$, the structure has a stretch-dominated behavior. Accordingly, a bending-dominated behavior corresponds to structures with $M< 0$. If the corresponding values from Table \ref{tab:struts_nodes} are used in Eq. \eqref{eq:maxwell}, only the popular \textsc{Octet} structure ($M = 0$) has a {stretch-dominated} behavior. The {BFCC} ($M = -9$), the \textsc{RhomOcta} ($M = -18$) and the \textsc{TrunOcta} ($M = -30$) all have a bending dominated. The drawback of these simple characteristics is, however, that large deformations of the lattice which include buckling and folding of layers are not weighted.


\begin{figure*} 
\centering
\resizebox{1\textwidth}{!}{
\begin{tikzpicture}
\draw (0,0) node[anchor=south west]{
\includegraphics[width=1\textwidth]{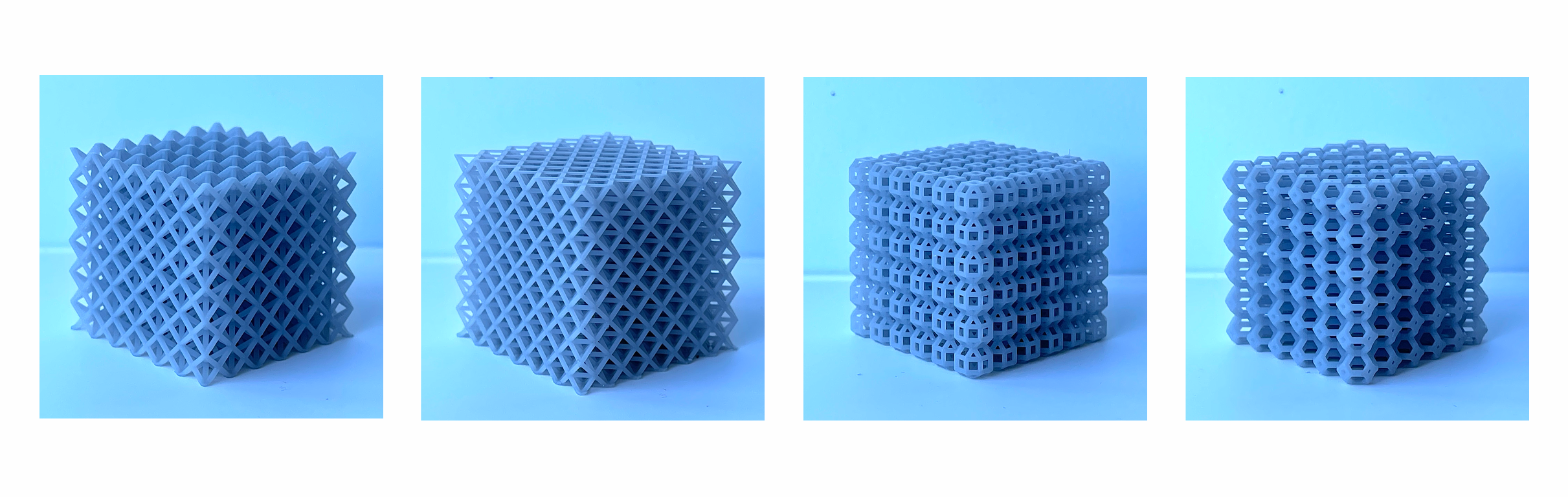}};
\draw(1.2,4) node[anchor=south west]{\small{\textsc{BFCC}}};
\draw(4.6,4) node[anchor=south west]{\small{\textsc{Octet}}};
\draw(7.2,4) node[anchor=south west]{\small{\textsc{RhomOcta}}};
\draw(10.5,4) node[anchor=south west]{\small{\textsc{TrunOcta}}};
\end{tikzpicture}}
\caption{Lattice-structure  specimens consisting of $6 \times  6 \times 6$ unit cells}
\label{printed_structures}
\end{figure*}
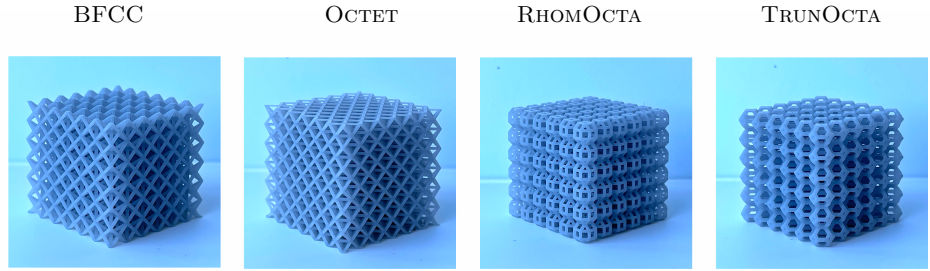

To make the results of our test comparable, the lattice structures are printed with a defined volume fraction $f_\text{V}$, which is the ratio of the actual strut volume $V$ to the reference volume $V_{\text{ref}}$ of the cube with the edge length $L$, 
\begin{align}
f_\text{V} =  \frac{V}{L^3} \label{eq:Vf} \,.
\end{align}
Here we investigate the two volume fractions of $f_\text{V}$ = 0.2 and $f_\text{V}$ = 0.3. All struts have the same thickness, and no additional mass is added to the nodes. The radius of the struts varies with the structure type and depends on the unit cell’s size. The edge length $L$ of the unit cell for the tests is 4.5\,mm for all sp types; thus, the struts’ radius is calibrated to the volume fraction, see Table \ref{tab:volume_fraction_lattice_structures}. The surface area of the unit cell for the different volume fractions of the lattice structures is also listed here. All specimens consists of $6 \times  6 \times 6$ unit cells, see Fig. \ref{printed_structures}.

\begin{figure*} 
\centering
\resizebox{1\textwidth}{!}{
\begin{tikzpicture}
\draw (0,0) node[anchor=south west]{
\includegraphics[width=1\textwidth]{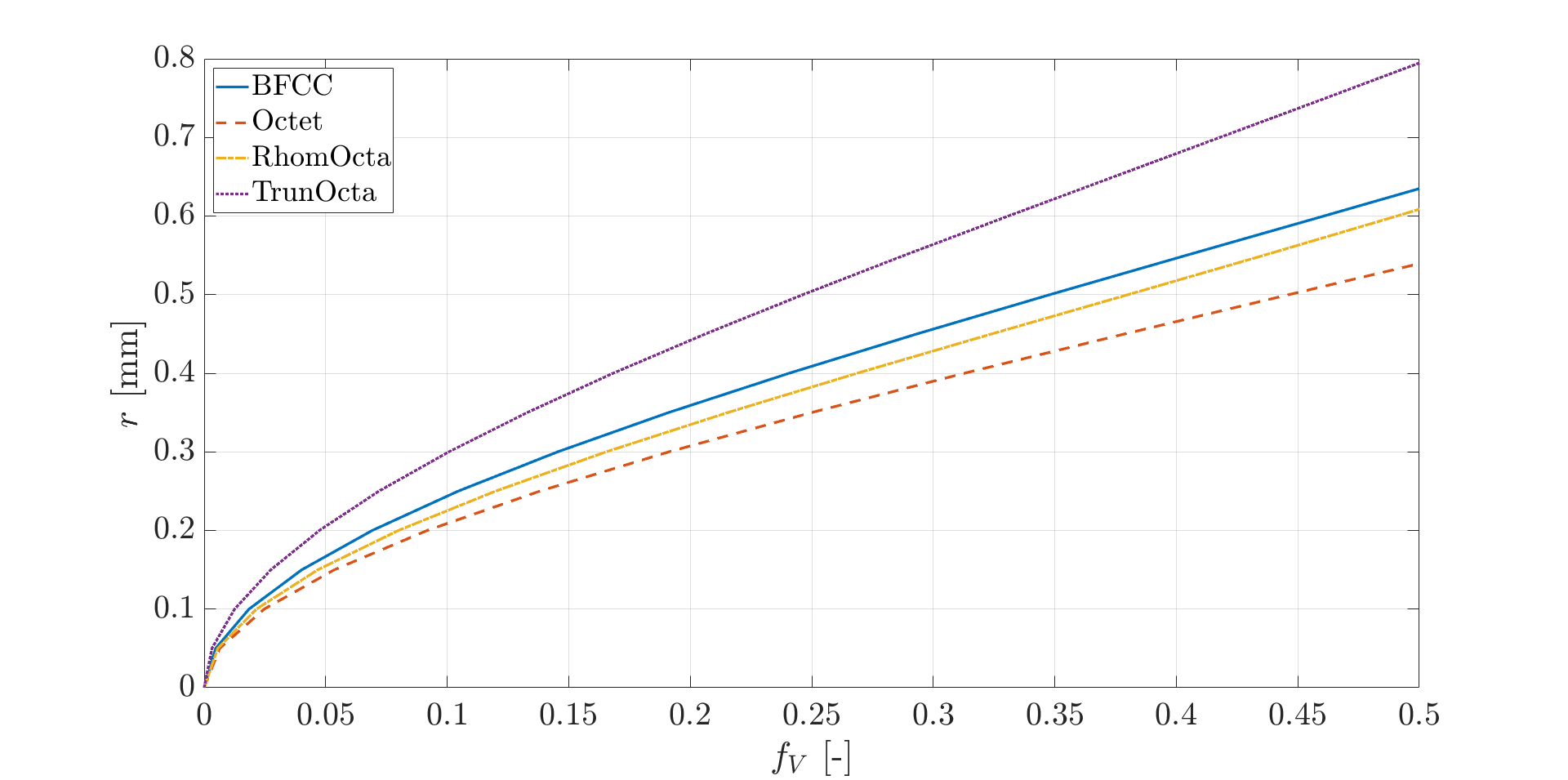}};
\draw(0.2,-0.3) node[anchor=south west]{};
\end{tikzpicture}}
\caption{Volume fraction $f_\text{V}$ vs. strut radius $r$ for different lattices}
\label{radius_surface_Vf}
\end{figure*}

We remark that the range of volume fraction for the comparison of different lattices is quite limited because the radius of the struts quickly becomes too thin to be fabricated (with reasonable accuracy) or too thick to model a truss-like open cell structure, cf. Fig.\,\ref{radius_surface_Vf} for an illustration.

\subsection{Printing process} \label{sec_printing_process}
%
Additive manufacturing offers numerous techniques for printing filigree structures. Very high-resolution and precise production is significant for the specimens used in this publication, as we use small unit cells. For this reason, all specimens are produced using Digital Light Processing technique. The printing process is a stereolithography (SLA)  where synthetic resin in liquid form is cured in layers by a light source. 
The high resolution of the panel installed in the printer allows a voxel size of up to 35\;$\upmu$m. 
With this resolution, the specimens can be printed with their different proportions without fusing the inner structure.

Due to the tiny unit cells, no support structures are required for the production of the specimens, which would be necessary for larger unit cells due to their higher weight, cf. \cite{bieler2023behavior}. This means that the post-processing of the specimens can be kept to a minimum. After printing, the specimens only need to be cleaned of excess liquid resin in a bath of isopropanol. After washing off the resin residue, the specimens are cured under UV light. All specimens were made from the material \textsc{Anycubic Tough 2.0}. This acrylate-based resin is characterized by a high toughness, but also by its resistance without being brittle. 
The resin can be printed with high precision and hardly shrinks.

%
\section{Experiments} \label{sec:chapter_Experiments}
%
For the uniaxial compression tests we used a universal tension/compression machine from \textsc{Hegewald \& Peschke}. The machine has a load cell of up to 2 kN. It  is equipped with compression plates between which the specimens are positioned and compressed. 

The lower compression plate is fixed, while the upper plate is attached to the movable crosshead. The crosshead speed has be adjusted to a constant speed of 1\,mm/s for all tests. The relatively small load cell allows very precise force values to be determined. This means the crosshead can initially be moved up to a preset force where  the specimen is slightly clamped. The corresponding force value was set to 0.05\,N here. The displacement of the traverse is then zeroed in order to obtain the correct values for the compression of the specimen for the data recording.
In the course of the experiment the specimen is compressed to  50\,\% of its initial height. 
This corresponds to a feed distance of the upper-pressure plate of 13.5\,mm.

\begin{figure} 
\centering
\resizebox{1\columnwidth}{!}{
\begin{tikzpicture}
\draw (0,0) node[anchor=south west]{
\includegraphics[width=1\textwidth]{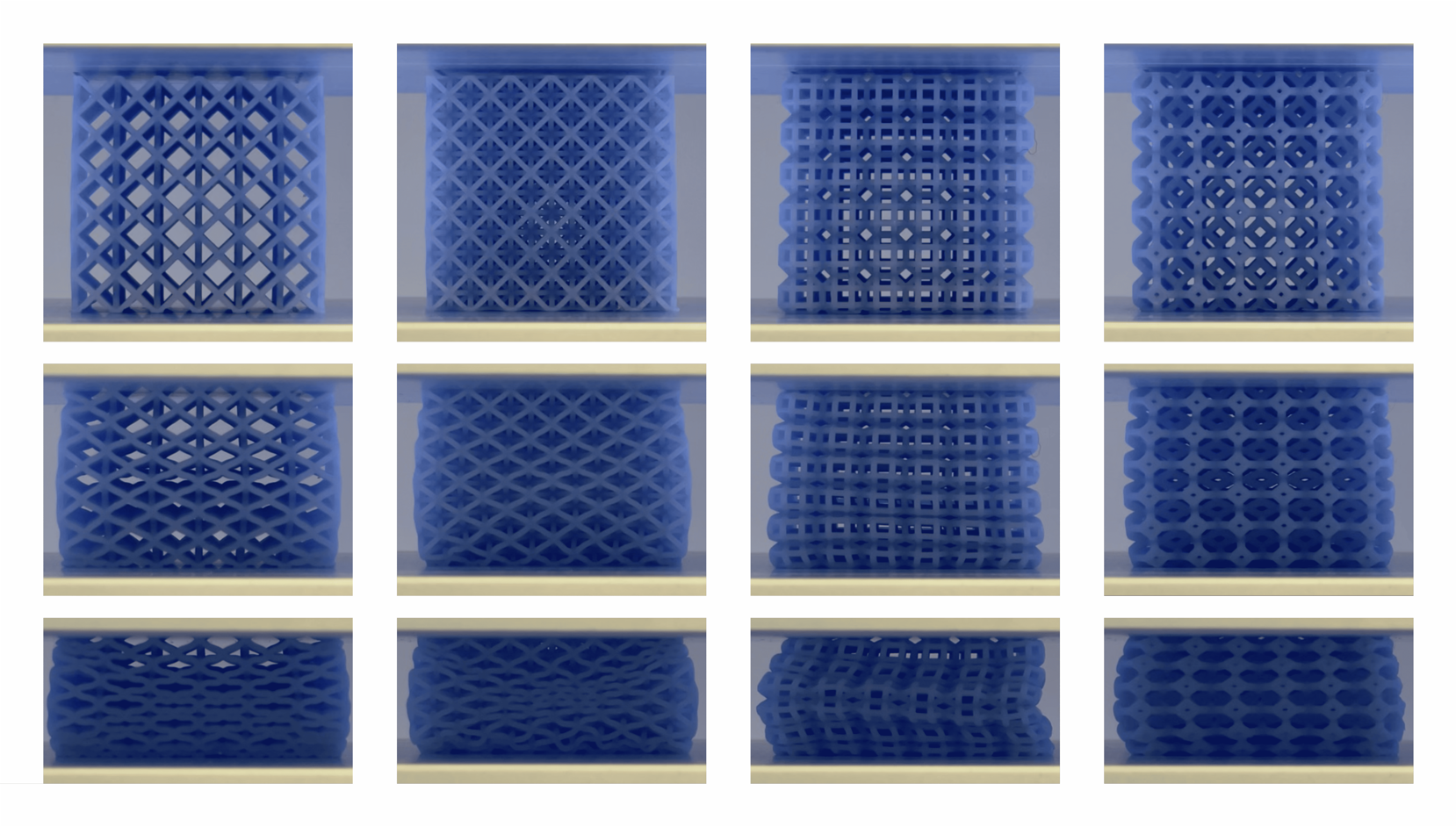}};
\draw(0,0.5) node[anchor=south west]{\begin{sideways}$\varepsilon=0.5$\end{sideways}};
\draw(0,2.4) node[anchor=south west]{\begin{sideways}$\varepsilon=0.25$\end{sideways}};
\draw(0,5.2) node[anchor=south west]{\begin{sideways}$\varepsilon=0$\end{sideways}};
\draw(10.4,7.2) node[anchor=south west]{\textsc{TrunOcta}};
\draw(7.1,7.2) node[anchor=south west]{\textsc{RhomOcta}};
\draw(4.4,7.2) node[anchor=south west]{\textsc{Octet}};
\draw(1.3,7.2) node[anchor=south west]{\textsc{BFCC}};
\end{tikzpicture}}
\caption{Investigated lattice structures with a volume fraction of $f_\text{V} = 0.2$  under $0$, $25$\,\% and $50$\,\% of compression }
\label{different_stages_of_deformation_Vf02}
\end{figure}
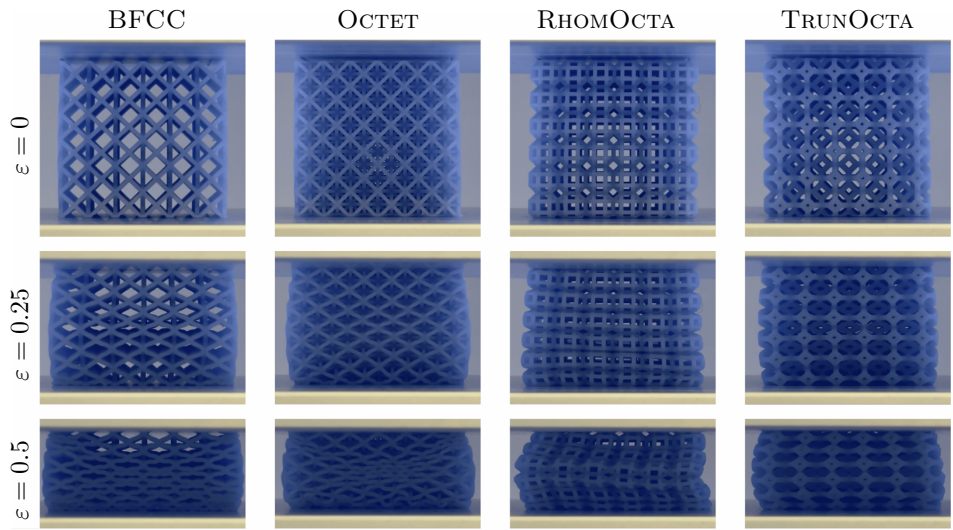

\begin{figure} 
\centering
\resizebox{1\columnwidth}{!}{
\begin{tikzpicture}
\draw (0,0) node[anchor=south west]{
\includegraphics[width=1\textwidth]{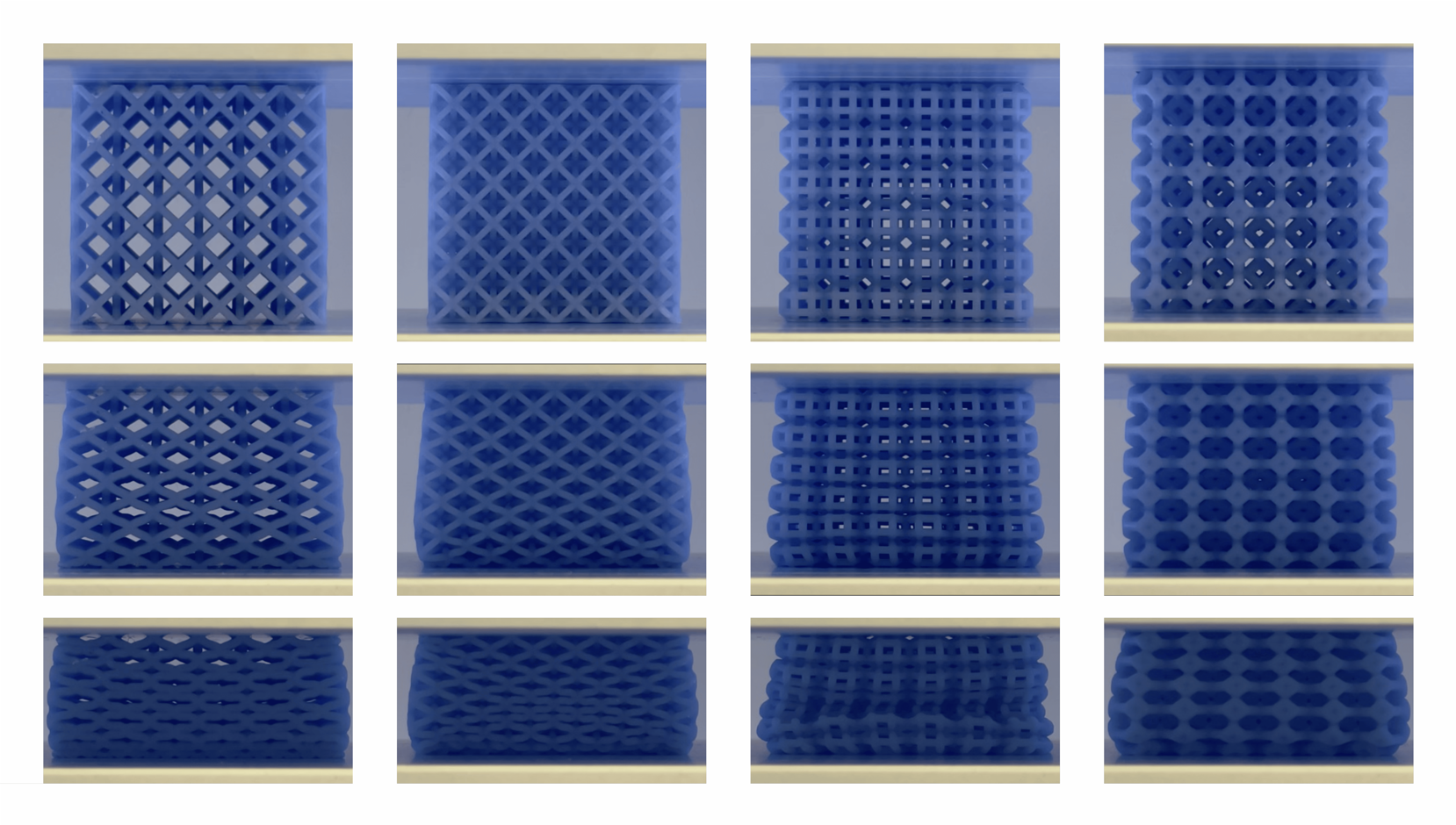}};
\draw(0,0.5) node[anchor=south west]{\begin{sideways}$\varepsilon=0.5$\end{sideways}};
\draw(0,2.4) node[anchor=south west]{\begin{sideways}$\varepsilon=0.25$\end{sideways}};
\draw(0,5.2) node[anchor=south west]{\begin{sideways}$\varepsilon=0$\end{sideways}};
\draw(10.4,7.2) node[anchor=south west]{\textsc{TrunOcta}};
\draw(7.1,7.2) node[anchor=south west]{\textsc{RhomOcta}};
\draw(4.4,7.2) node[anchor=south west]{\textsc{Octet}};
\draw(1.3,7.2) node[anchor=south west]{\textsc{BFCC}};
\end{tikzpicture}}
\caption{Investigated lattice structures with a volume fraction of $f_\text{V} = 0.3$  under $0$, $25$\,\% and $50$\,\% of compression }
\label{different_stages_of_deformation_Vf03}
\end{figure}

\begin{figure} 
\centering
\resizebox{1\columnwidth}{!}{
\begin{tikzpicture}
\draw (0,0) node[anchor=south west]{
\includegraphics[width=1\textwidth]{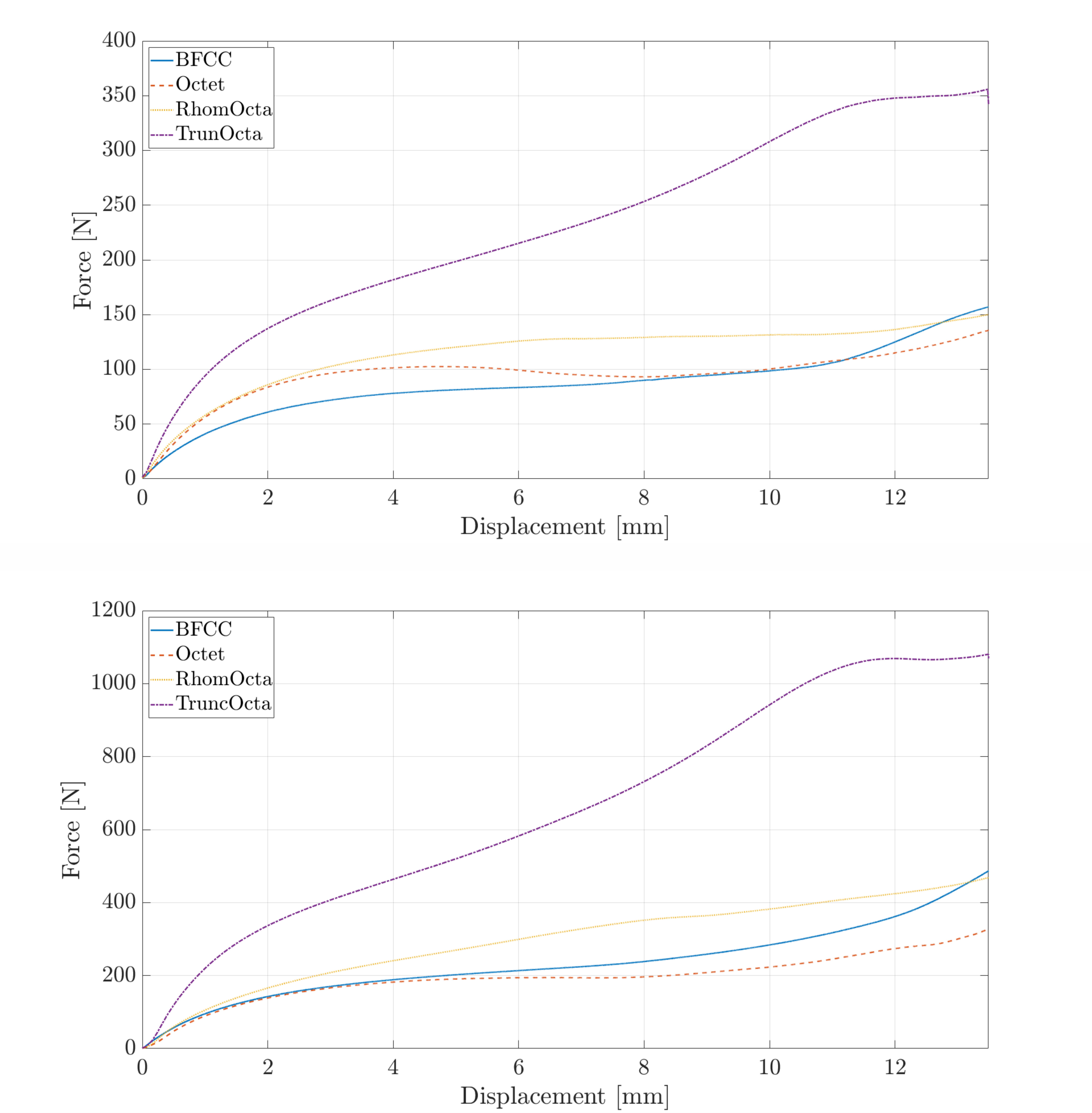}};
\draw(0,0) node[anchor=south west]{};
\end{tikzpicture}}
\caption{Recorded force-displacement curve for the lattice structures under compression with $f_\text{V}$ = 0.2 (top) and $f_\text{V}$ = 0.3 (bottom)}
\label{force_displacement}
\end{figure}

The snapshots in Fig. \ref{different_stages_of_deformation_Vf02} and \ref{different_stages_of_deformation_Vf03} shows different stages of compression for the  structures of $f_\text{V} = 0.2$ and $f_\text{V} = 0.3$. We observe that the deformation is quite uniform before buckling occurs. 
The corresponding force-over-displacement curves are shown in Fig.\,\ref{force_displacement} and pronounces the different responses for the different types of lattices. It appears that the \textsc{TrunOcta} structure is significantly stiffer than the other three lattices of same volume fraction. 

It is interesting to note the \textsc{Octet} structure, which at the first glance is quite similar to the  \textsc{Octahedron}, reacts very weak to the applied force and, accordingly, undergoes larger deformation than the other 
lattices. 

For all structure, the three phases of the  force-over-displacement curve can nicely be seen. 
A strong increase at the beginning of the compression corresponds to phase I,  followed by a pronounced phase-II force plateau between 15\% and 40 \% compression. If the single layers are pressed together, i.e. the layers come into contact with each other, the structure looses its open-cell character and which in turn causes the force to show a rapid phase-III increase. 

The \textsc{TrunOcta} shows a much steeper phase-I increase in force for both investigated volume fractions. 
Due to folding of individual layers of the specimens, a more or less pronounced  phase-II force plateau is formed during the compression.  This plateau is mainly observed for the structures \textsc{BFCC} and \textsc{Octet} for both volume fractions. The \textsc{RhomOcta} structure shows a later onset of the plateau for $f_\text{V}$ = 0.2, whereas for  $f_\text{V}$ = 0.3, there is no longer an apparent  phase-II plateau. It is noticeable that the \textsc{TrunOcta} structure behaves stiff, and a typical phase-II plateau is basically absent. 
After the phase-II, the force curve rises again because of the phase-III of collapsed layers.  This can be seen in Fig. \ref{force_displacement} for the structures \textsc{BFCC} and \textsc{Octet}, whereas the \textsc{TrunOcta} does not yet show such a strong compaction. 

In Fig.~\ref{different_stages_of_deformation_Vf02} and \ref{different_stages_of_deformation_Vf03}, we also see that the \textsc{BFCC} and \textsc{Octet} structures deform symmetrically up to $\varepsilon = 0.5$. 
One reason may be that both have nodes in the corners and the center of the unit cell, so they react better to the orthogonal force. An asymmetric deformation state is identifiable for the two convex structures with no nodes in the center of the unit cell. 
Here, the structure \textsc{TrunOcta}   has 25\,\% thicker struts than \textsc{RhomOcta} at the same volume fraction. This contributes to the bending stiffness of the structure. 


We emphasize that despite of severe deformation, no externally visible fracture was found in any of the specimens. All specimens returned to their initial state after a recovery time  between 15 and 45 minutes. 


\subsection{Energy absorption}
\begin{figure} 
\centering
\resizebox{1\columnwidth}{!}{
\begin{tikzpicture}
\draw (0,0) node[anchor=south west]{
\includegraphics[width=1\textwidth]{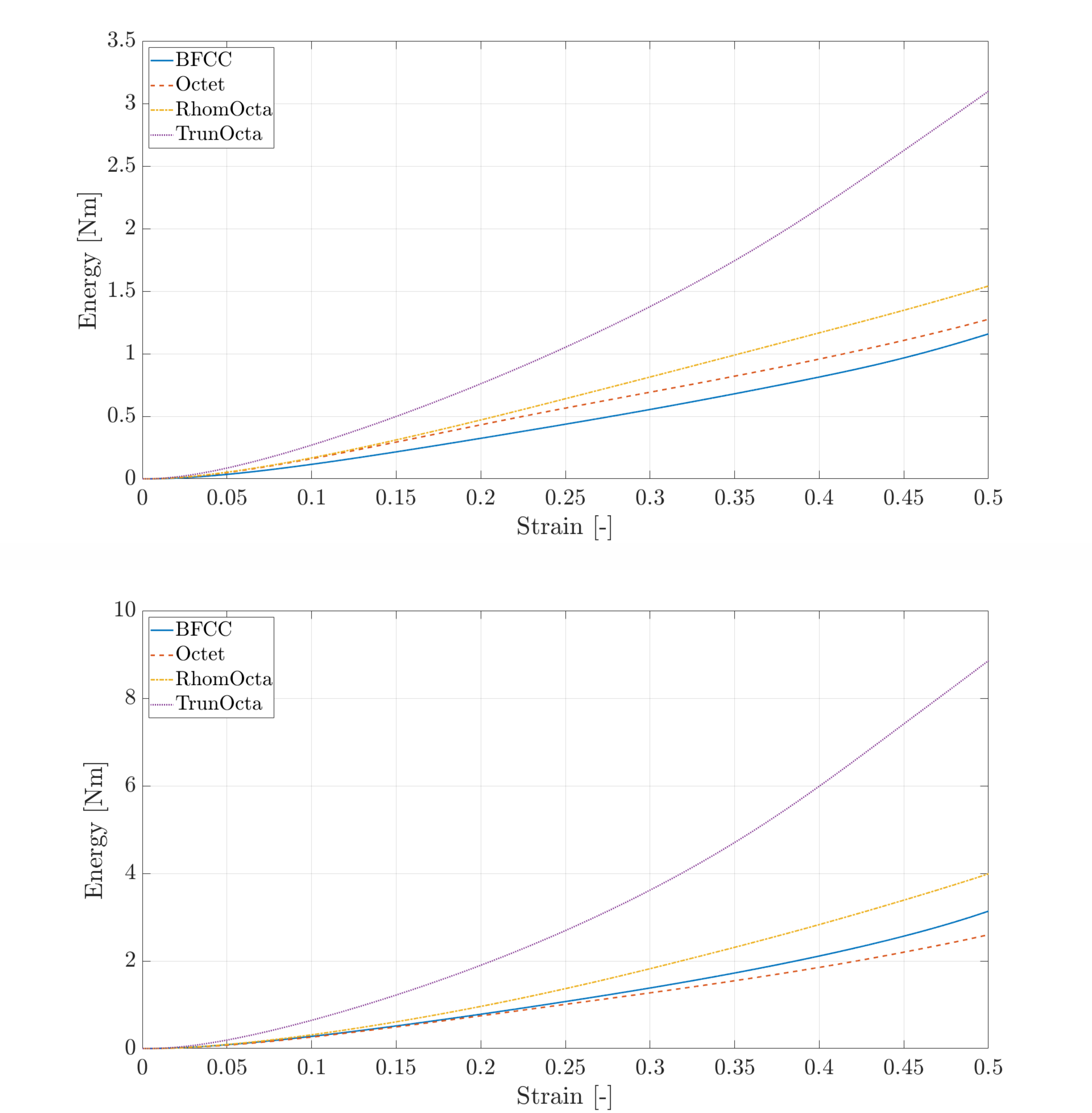}};
\draw(0,0) node[anchor=south west]{ };
\end{tikzpicture}}
\caption{Energy absorption for different states of deformation during the compression test for specimen with volume fraction of $f_\text{V} = 0.2$ (top) and $f_\text{V} = 0.3$ (bottom)}
\label{fig:energy_absorption_vf02_03}
\end{figure}

Goal of our investigation was to determine the energy 'absorbed' during the (reversible) deformation, i.e., the energy the lattice structures stores during compression. To this end, the energy applied to the specimen is calculated from the measured force $F$ and the displacement $d$ of the machine's  upper crosshead,
\begin{align} \label{eq:energy_traverse}
W = 	\int_{0}^{d} F \, \mathrm{d}x \,.
\end{align}
Fig. \ref{fig:energy_absorption_vf02_03} shows the energy absorption of the lattice structures over the strain,
\begin{align} \label{eq:strain_traverse}
\varepsilon = 	\frac{d}{6\,L} 
\end{align}
Additionally, we define the (mass) specific energy absorption (SEA) as   the quotient of the stored energy and the specimen's mass, 
\begin{align}
\text{SEA} = 	\frac{W}{m} \label{eq:SEA}
\end{align}
The mass $m$ of the specimen is calculated from the density of the resin $\rho$, the volume fraction $f_\text{V}$, the unit cell's length $L$ and the number of cells, 
\begin{align}\label{eq:mass}
m = 216 \, \rho \, f_\text{V} \, L^3   \,.
\end{align}

\begin{figure} 
\centering
\resizebox{1\columnwidth}{!}{
\begin{tikzpicture}
\draw (0,0) node[anchor=south west]{
\includegraphics[width=1\textwidth]{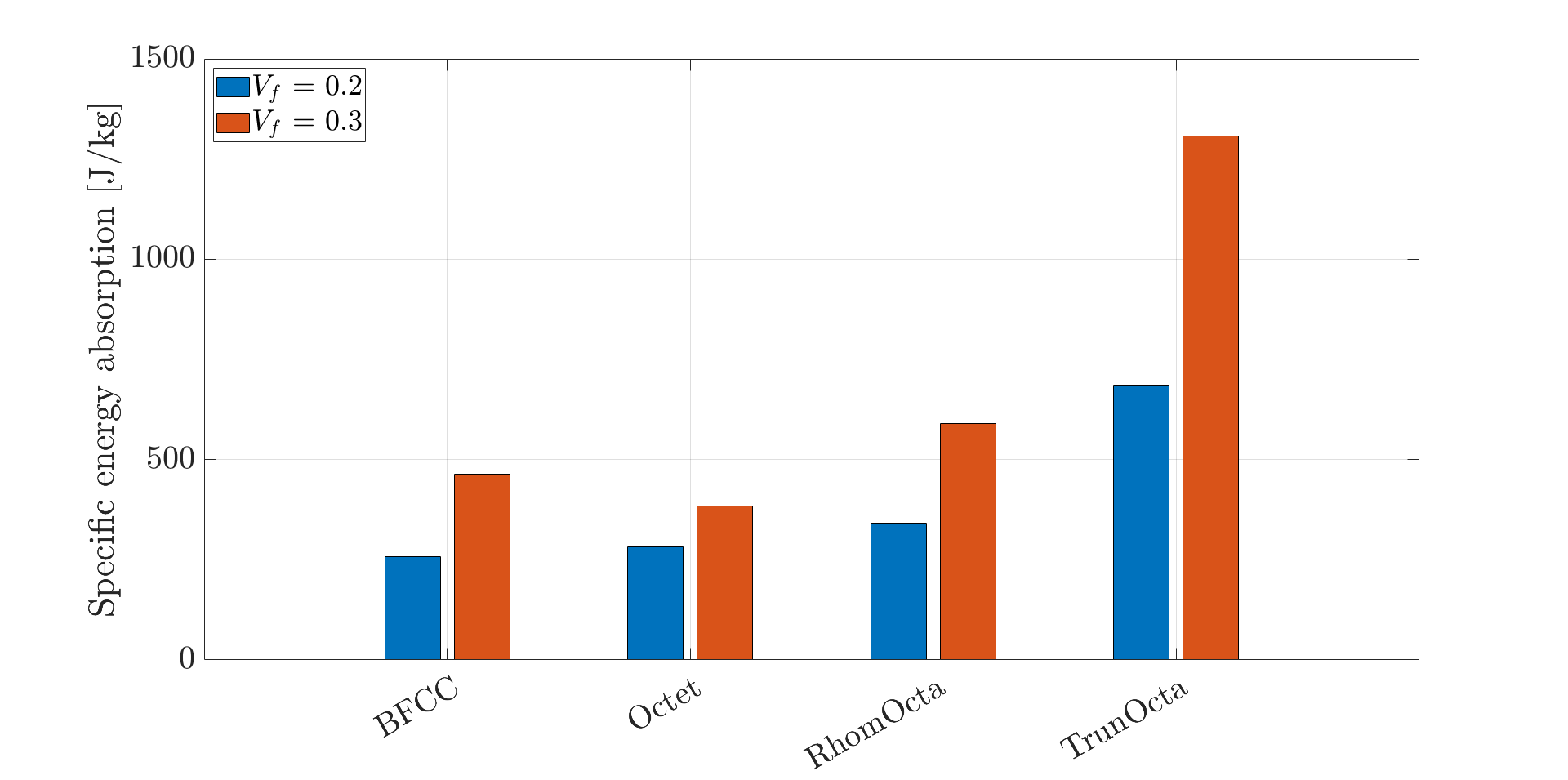}};
\draw(0,0) node[anchor=south west]{ };
\end{tikzpicture}}
\caption{Specific energy absorption for specimen compressed by $\varepsilon = 0.5$}
\label{fig:specific_energy_absorption}
\end{figure}

Naturally, the energy absorption relates to the forces required for compression. The \textsc{TrunOcta} shows a significantly higher SEA than the other structures, see Fig. \ref{fig:specific_energy_absorption}. In contrast, the \textsc{BFCC} and the \textsc{Octet} structures behave similarly. The \textsc{RhomOcta} structure absorbs somewhat more energy, whereby the tests have shown that an asymmetrical folding of the individual layers can occur which may lead to greater energy absorption. 

Considering the SEA, we clearly see the \textsc{TrunOcta} lattice to perform best. This finding has significant practical implications, as it suggests that the \textsc {TrunOcta} lattice could be a promising candidate for applications requiring high energy absorption in lattice structures.

One topic left out of this work is possible imperfections in the printed structures. The authors \cite{glaesener2023predicting} have also dealt with this topic by limiting the type of imperfection.

%

\section{Numerical simulation of the compression test} \label{chapter_Simulation}
%
In order to understand the mechanisms of energy absorption in the different lattice types, we reproduce the compression tests numerically. 

\begin{figure} 
\centering
\resizebox{1\columnwidth}{!}{
\begin{tikzpicture}
\draw (0,0) node[anchor=south west]{
\includegraphics[width=1\textwidth]{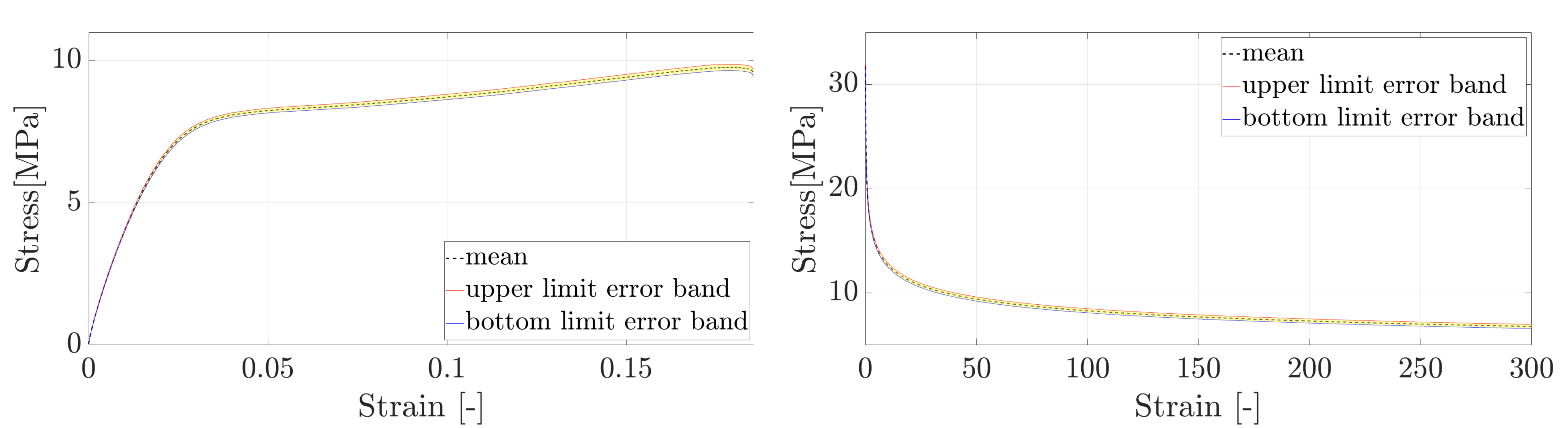}};
\draw(0,0) node[anchor=south west]{ };
\end{tikzpicture}}
\caption{Tension test data (left) and relaxation test data (right) of a dog bone specimens with an error band based on a confidence interval of 95 \%}
\label{fig:tension_test_dog_bone}
\end{figure}

We started determining the material parameters required for the FEA. For this purpose, were carried out tension tests and relaxation tests performed on our tensile test machine. For both tests the specimen geometry was selected as specimen body type 5A from DIN EN ISO 527-2. In the tensile test, the specimens were stretched until failure at a speed of 0.5 mm/min, and the stress-strain diagram was recorded, cf. Fig. \ref{fig:tension_test_dog_bone} (left).

In the relaxation test the specimens with a nominal length of $L_0$ = 50\,mm were abruptly stretched by $2.5$\,mm at a feed speed of 300\,mm/min. This corresponds to a strain of 5\,\%. The elongation was then maintained, and the force decrease over time was measured for 300\,s. The corresponding stress-time curves were used to determine a viscoelastic relaxation model using a Prony series, i.e.
\begin{align}\label{eq:Prony}
  E(t) = E_{\infty} \left( 1 +\sum_{i=1}^N g_i \exp{(-t/\tau_i)}\right) \,,
\end{align}
see also Fig.\,\ref{fig:tension_test_dog_bone} (right). Here $E_\infty$ is the elastic modulus at the relaxed state, $g_i$ and $\tau_i$ are relaxation module in dimensionless from and the corresponding relaxation times. The identified Prony coefficients are in listed in Table \ref{tab:prony} together with the other material parameter of \textsc{Anycubic Tough} material. 

%

\begin{table*} 
\caption{Elastic modulus and viscoelastic Prony coefficients of printed \textsc{Anycubic Tough} material determined experimentally; according to a Poisson's ratio is $\nu=0.37$ and a density of the material of 1.15 g/cm$^3$}
\label{tab:prony} 
\begin{tabularx}{\textwidth}{XXXX} \toprule
\noalign{\smallskip}
$E$  & Prony coefficients & $g_i$ & $\tau_i$\\
\noalign{\smallskip}
\hline
\noalign{\smallskip}
 & i=1 & 0.35399 & 0.0303178\\
\noalign{\smallskip}
393 MPa & i=2 & 0.27197 & 5.0618\\
\noalign{\smallskip}
& i=3 & 0.16004 & 84.047\\
\noalign{\smallskip}
\hline
\end{tabularx}
\end{table*}

\begin{figure} 
\centering
\resizebox{0.7\columnwidth}{!}{
\begin{tikzpicture}
\draw (0,0) node[anchor=south west]{
\includegraphics[width=1\textwidth]{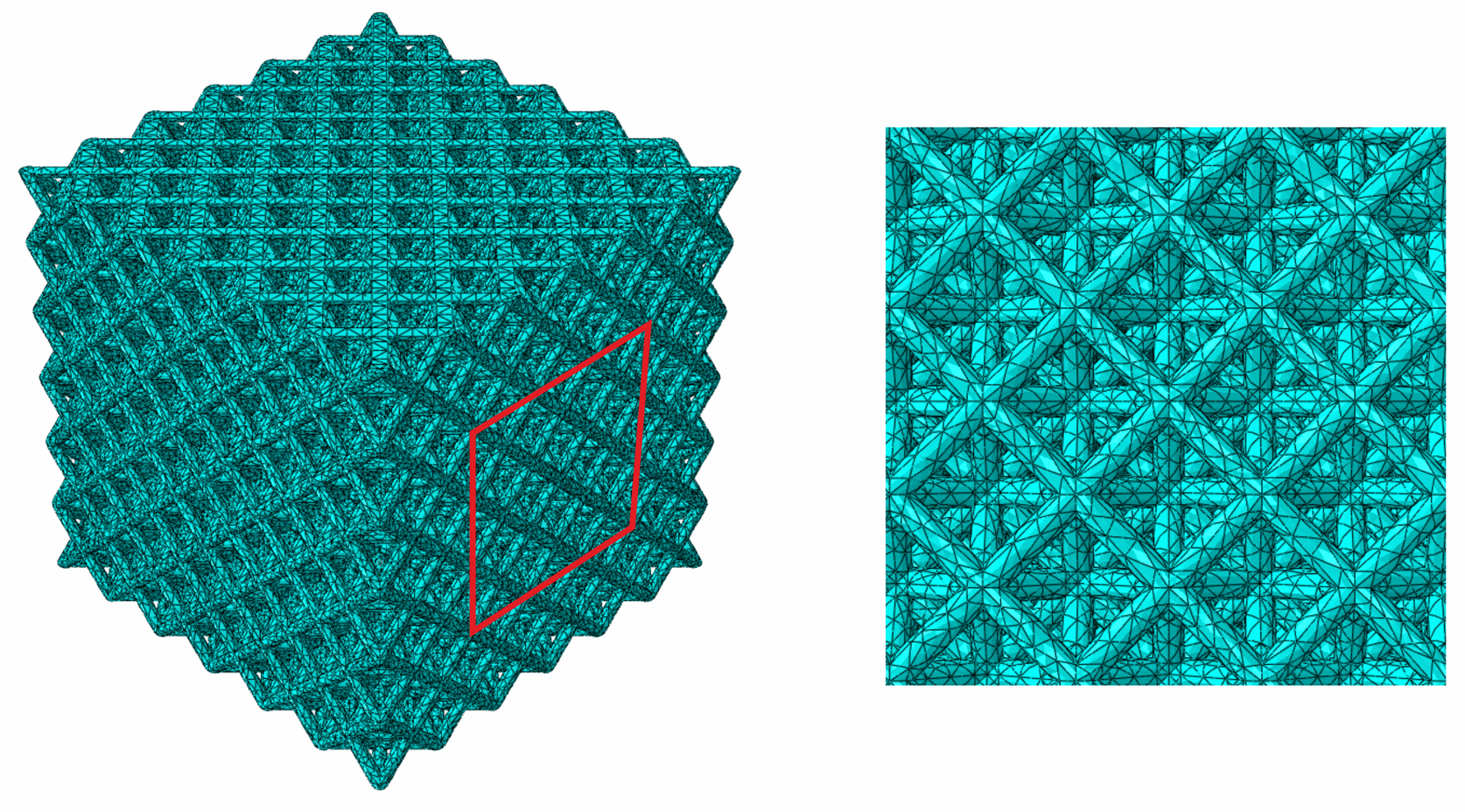}};
\draw(0,0) node[anchor=south west]{ };
\end{tikzpicture}}
\caption{FEA mesh density for the \textsc{Octet} structure}
\label{fig:mesh}
\end{figure}

The FEAs were computed with \textsc{Abaqus}, details of the finite element mesh are shown in Fig.~\ref{fig:mesh}. However, when facing the challenges of lattice-strut buckling and the subsequent collapse of cell layers, a non-linear static FEA is not sufficient. The symmetric arrangement makes it too deterministic to find the points of instability. Thus, we employ dynamic FEAs with \textsc{ABAQUS/Explicit} to simulate the static test using dynamic relaxation.

\begin{figure} 
\centering
\resizebox{1\columnwidth}{!}{
\begin{tikzpicture}
\draw (0,0) node[anchor=south west]{
\includegraphics[width=1\textwidth]{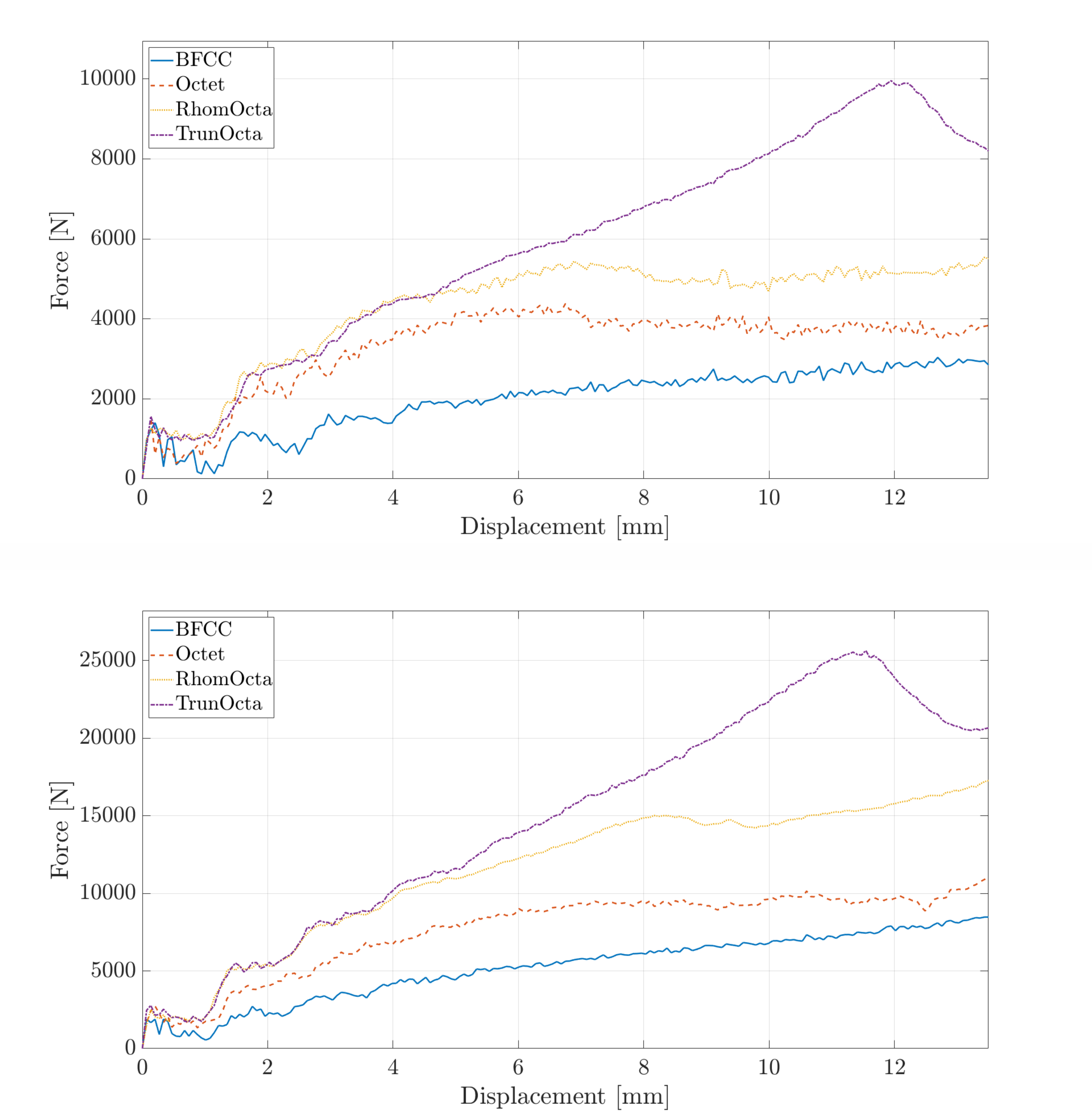}};
\draw(0,0) node[anchor=south west]{};
\end{tikzpicture}}
\caption{Force-displacement curve from the simulations carried out with evaluation of the reference point on the upper moving plate}
\label{fig:FEA:force_displacement}
\end{figure}

The two compression plates were created as \textsc{Rigid} bodies, which do not allow any deformation. The boundary conditions of the lower plate were fixed in all directions. The boundary conditions of the upper plate only allow movements in the $z$-direction. The upper plate is moved via a reference point created on its surface. The specimen is arranged between the plates and free to slide. For the contact conditions, \textsc{Self Contact} was selected as \textsc{General Contact}. This also applies to the self-contact conditions within the specimen.  For the \textsc{Interaction Properties}, a frictional behavior with a friction coefficient of 0.15 was assumed. 
Due to their complex structure, the specimens were meshed with ten-node modified quadratic tetrahedron elements (C3D10M). The number of elements for the different lattice structures are listed in Table \ref{tab:elements}. The specimens were compressed up to  50\,\%, as in the experiments, which corresponds to 13.5\,mm, see Fig.\ref{fig:FEA:force_displacement}.

\begin{figure}
\centering
\resizebox{1\columnwidth}{!}{
\begin{tikzpicture}
\draw (0,0) node[anchor=south west]{
\includegraphics[width=1\textwidth]{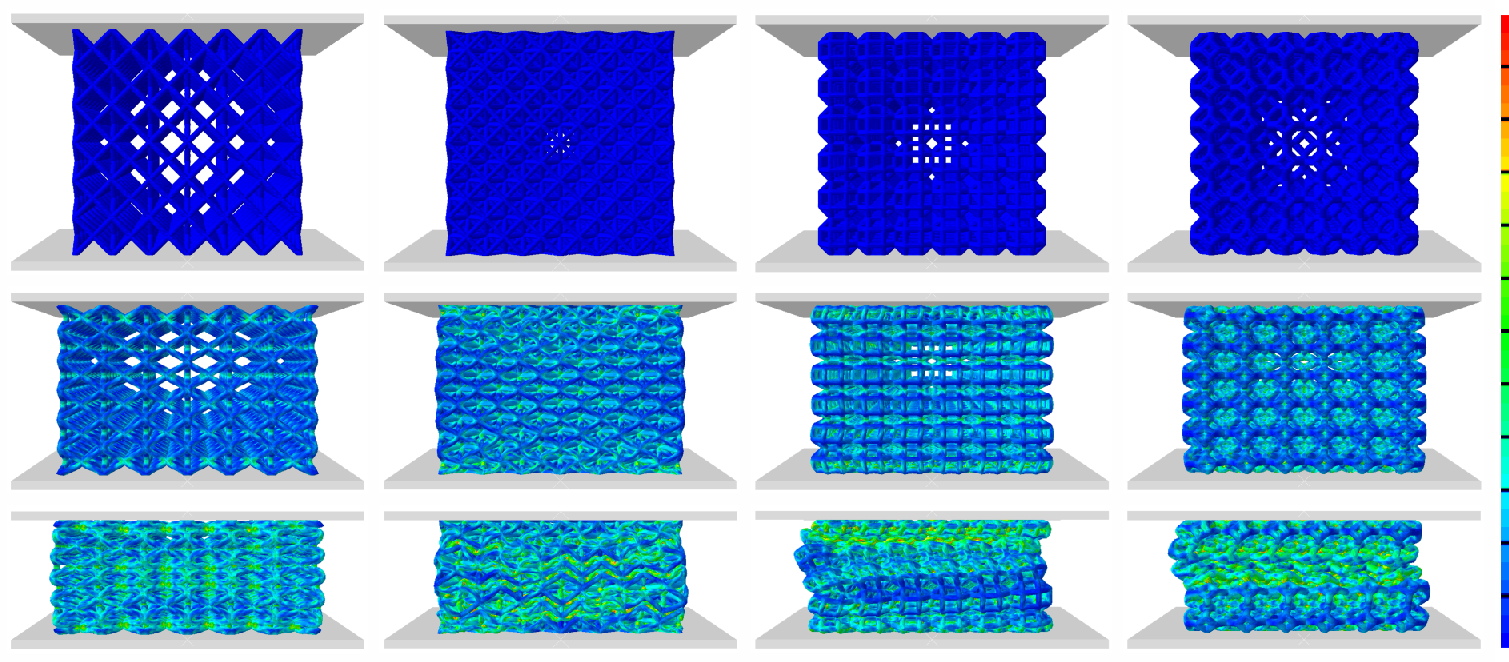}};
\draw(-0.2,0.15) node[anchor=south west]{\begin{sideways}$\varepsilon=0.5$\end{sideways}};
\draw(-0.2,1.7) node[anchor=south west]{\begin{sideways}$\varepsilon=0.25$\end{sideways}};
\draw(-0.2,4.05) node[anchor=south west]{\begin{sideways}$\varepsilon=0$\end{sideways}};
\draw(10.3,6) node[anchor=south west]{\textsc{TrunOcta}};
\draw(7.1,6) node[anchor=south west]{\textsc{RhomOcta}};
\draw(4.4,6) node[anchor=south west]{\textsc{Octet}};
\draw(1.2,6) node[anchor=south west]{\textsc{BFCC}};
\draw(13.2,1.8) node[anchor=south west]{\begin{sideways}S, Mises (MPa)\end{sideways}};
\draw(13.2,0.2) node[anchor=south west]{min};
\draw(13.2,5.4) node[anchor=south west]{max};
\end{tikzpicture}}
\caption{Simulation of the compression test of the investigated lattice structures with a volume fraction of $f_\text{V} = 0.2$ at different stages of deformation}
\label{deformation_abaqus_vf02}
\end{figure}

\begin{figure} 
\centering
\resizebox{1\columnwidth}{!}{
\begin{tikzpicture}
\draw (0,0) node[anchor=south west]{
\includegraphics[width=1\textwidth]{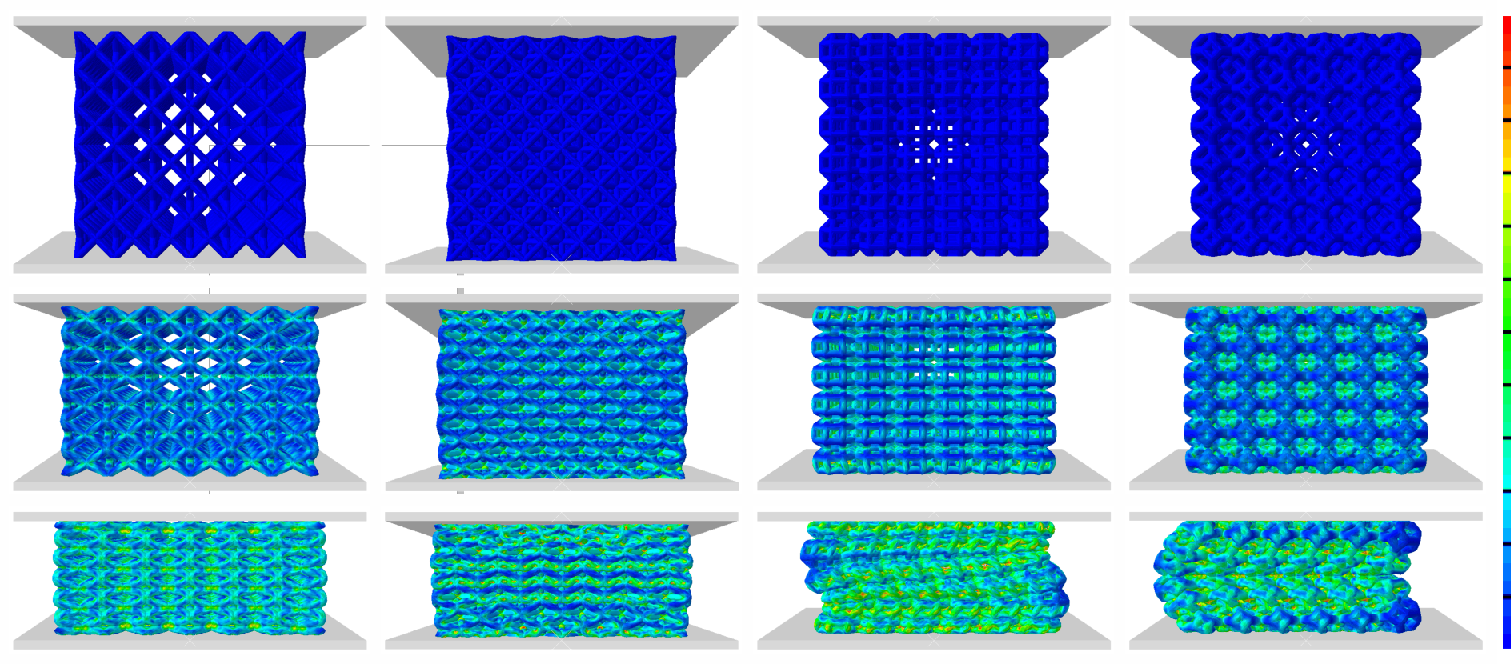}};
\draw(-0.2,0.15) node[anchor=south west]{\begin{sideways}$\varepsilon=0.5$\end{sideways}};
\draw(-0.2,1.7) node[anchor=south west]{\begin{sideways}$\varepsilon=0.25$\end{sideways}};
\draw(-0.2,4.05) node[anchor=south west]{\begin{sideways}$\varepsilon=0$\end{sideways}};
\draw(10.3,6) node[anchor=south west]{\textsc{TrunOcta}};
\draw(7.1,6) node[anchor=south west]{\textsc{RhomOcta}};
\draw(4.4,6) node[anchor=south west]{\textsc{Octet}};
\draw(1.2,6) node[anchor=south west]{\textsc{BFCC}};
\draw(13.2,1.8) node[anchor=south west]{\begin{sideways}S, Mises (MPa)\end{sideways}};
\draw(13.2,0.2) node[anchor=south west]{min};
\draw(13.2,5.4) node[anchor=south west]{max};
\end{tikzpicture}}
\caption{Simulation of the compression test of the investigated lattice structures with a volume fraction of $f_\text{V} = 0.3$ at different stages of deformation}
\label{deformation_abaqus_vf03}
\end{figure}

\begin{figure} 
\centering
\resizebox{1\columnwidth}{!}{
\begin{tikzpicture}
\draw (0,0) node[anchor=south west]{
\includegraphics[width=1\textwidth]{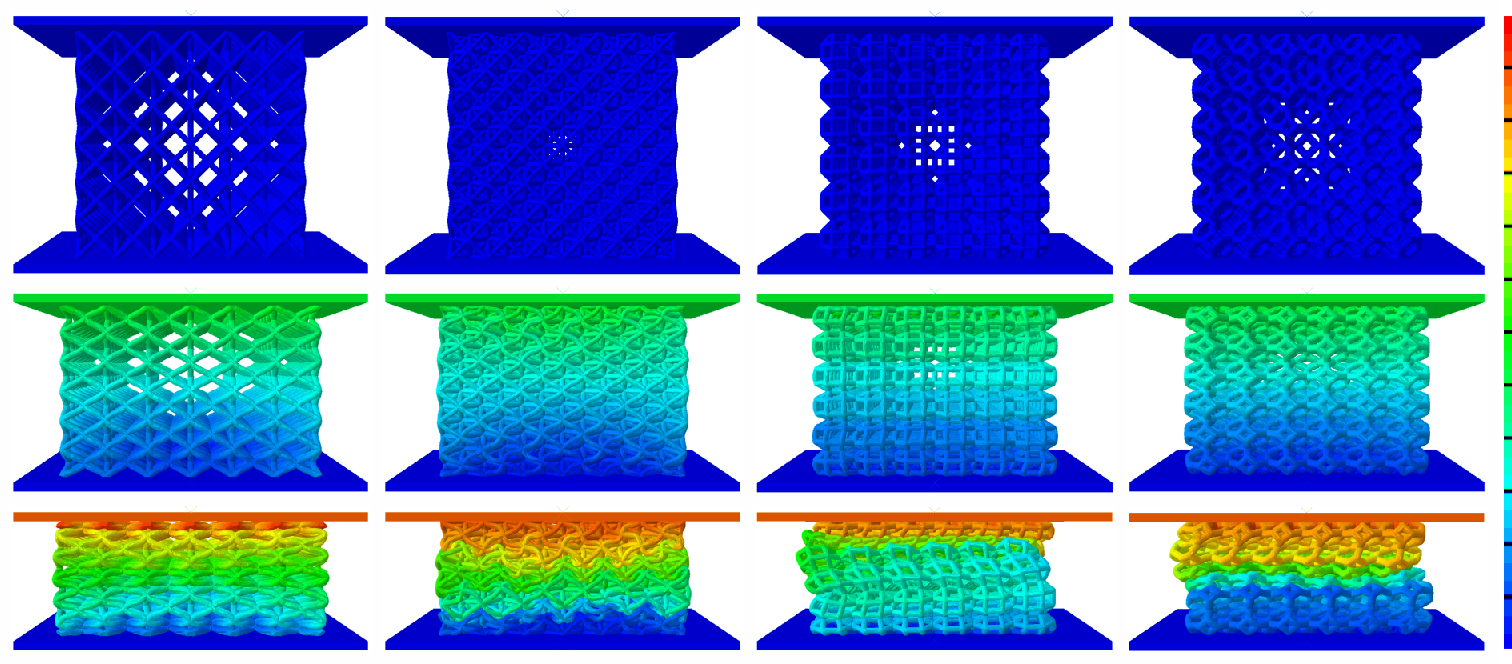}};
\draw(-0.2,0.15) node[anchor=south west]{\begin{sideways}$\varepsilon=0.5$\end{sideways}};
\draw(-0.2,1.7) node[anchor=south west]{\begin{sideways}$\varepsilon=0.25$\end{sideways}};
\draw(-0.2,4.05) node[anchor=south west]{\begin{sideways}$\varepsilon=0$\end{sideways}};
\draw(10.3,6) node[anchor=south west]{\textsc{TrunOcta}};
\draw(7.1,6) node[anchor=south west]{\textsc{RhomOcta}};
\draw(4.4,6) node[anchor=south west]{\textsc{Octet}};
\draw(1.2,6) node[anchor=south west]{\textsc{BFCC}};
\draw(13.2,1.8) node[anchor=south west]{\begin{sideways}U (mm)\end{sideways}};
\draw(13.2,0.2) node[anchor=south west]{0};
\draw(13.2,5.4) node[anchor=south west]{15};
\end{tikzpicture}}
\caption{Simulation of the compression test of the investigated lattice structures with a volume fraction of $f_\text{V} = 0.2$ at different stages of deformation}
\label{deformation_abaqus_vf02_2}
\end{figure}

\begin{figure} 
\centering
\resizebox{1\columnwidth}{!}{
\begin{tikzpicture}
\draw (0,0) node[anchor=south west]{
\includegraphics[width=1\textwidth]{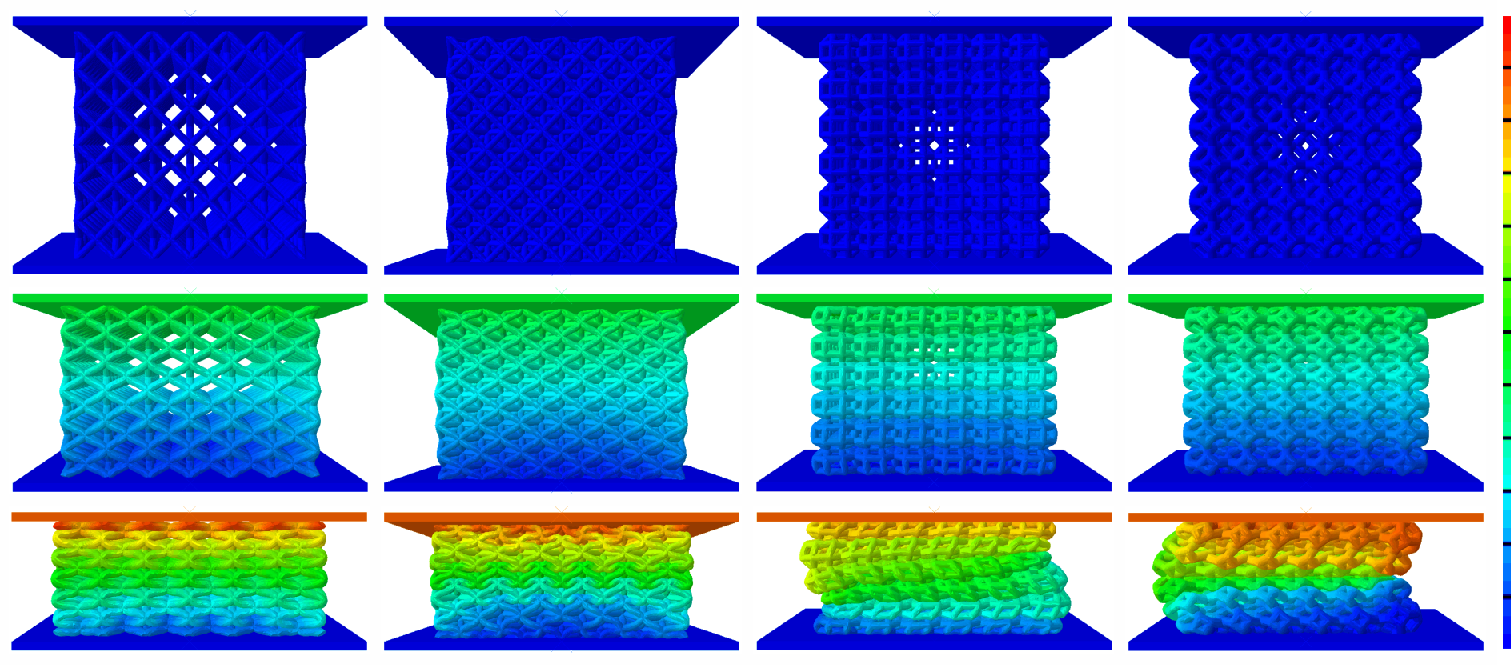}};
\draw(-0.2,0.15) node[anchor=south west]{\begin{sideways}$\varepsilon=0.5$\end{sideways}};
\draw(-0.2,1.7) node[anchor=south west]{\begin{sideways}$\varepsilon=0.25$\end{sideways}};
\draw(-0.2,4.05) node[anchor=south west]{\begin{sideways}$\varepsilon=0$\end{sideways}};
\draw(10.3,6) node[anchor=south west]{\textsc{TrunOcta}};
\draw(7.1,6) node[anchor=south west]{\textsc{RhomOcta}};
\draw(4.4,6) node[anchor=south west]{\textsc{Octet}};
\draw(1.2,6) node[anchor=south west]{\textsc{BFCC}};
\draw(13.2,1.8) node[anchor=south west]{\begin{sideways}U (mm)\end{sideways}};
\draw(13.2,0.2) node[anchor=south west]{0};
\draw(13.2,5.4) node[anchor=south west]{15};
\end{tikzpicture}}
\caption{Simulation of the compression test of the investigated lattice structures with a volume fraction of $f_\text{V} = 0.3$ at different stages of deformation}
\label{deformation_abaqus_vf03_2}
\end{figure}

We start comparing the deformation of the different lattice structures as shown in Fig.~\ref{deformation_abaqus_vf02} and Fig.~\ref{deformation_abaqus_vf02_2} for $f_\text{V}=0.2$ and \ref{deformation_abaqus_vf03} and Fig.~\ref{deformation_abaqus_vf03_2} for $f_\text{V} =0.3$. For both volume fractions the uniform deformation of the structures of \textsc{Octet} and \textsc{BFCC}, as well as the buckling structure of \textsc{RhomOcta} specimens is nicely reflected in the FEA. 

In contrast, the uniform deformation from the experiments with the \textsc{TrunOcta} structure does not exactly match the results of the simulation. Here we observe instabilities in the FEA which we do not see in the experiments. 

\begin{figure} 
\centering
\resizebox{1\columnwidth}{!}{
\begin{tikzpicture}
\draw (0,0) node[anchor=south west]{
\includegraphics[width=1\textwidth]{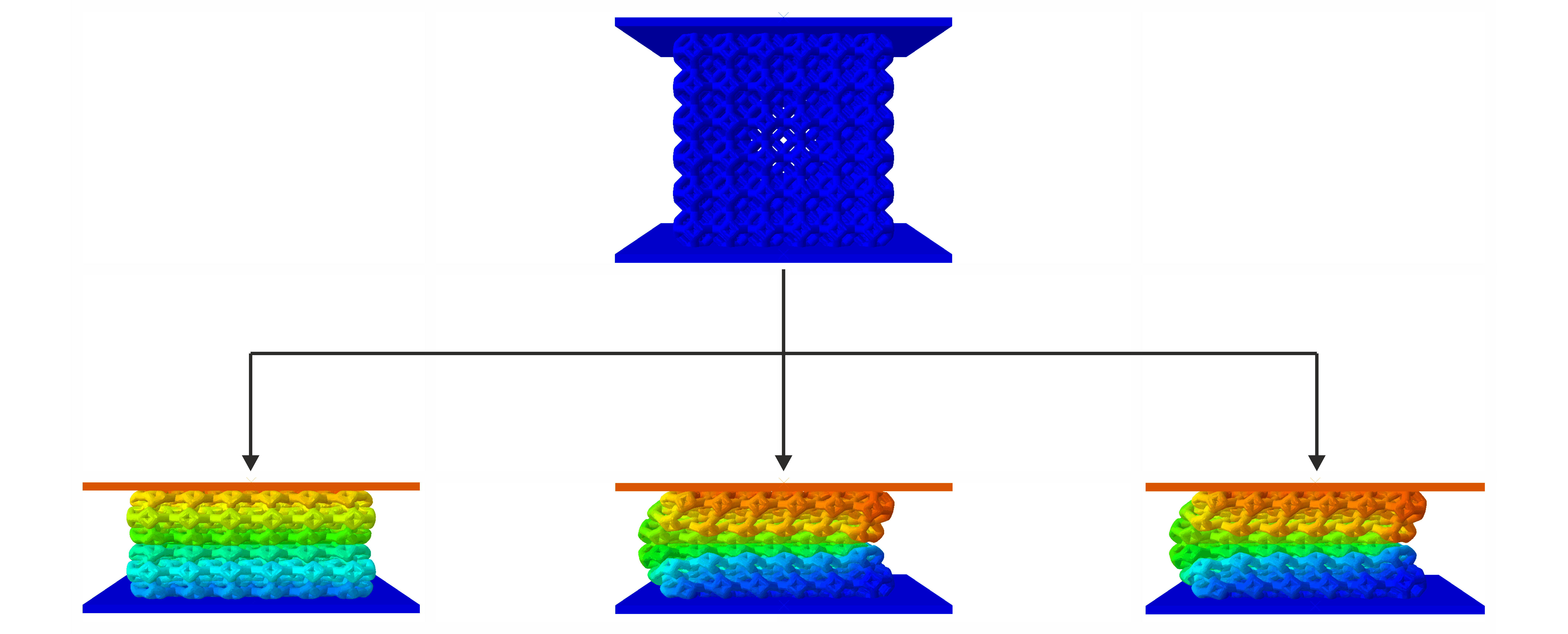}};
\draw(-0.2,0.15) node[anchor=south west]{\begin{sideways}$\varepsilon=0.5$\end{sideways}};
\draw(5.6,5.3) node[anchor=south west]{\textsc{TrunOcta}};
\draw(1.1,-0.3) node[anchor=south west]{$E = 200$ MPa};
\draw(4.6,-0.3) node[anchor=south west]{$E = 400$ MPa (measured)};
\draw(10,-0.3) node[anchor=south west]{$E = 800$ MPa};
\draw(0,-0.5) node[anchor=south west]{ };
\end{tikzpicture}}
\caption{Comparison of FEA of the \textsc{TrunOcta} structure ($f_\text{V} = 0.3$) with different moduli of elasticity}
\label{fig:deformation_abaqus_vf03_2}
\end{figure}

In some sense, the deformation observed in the FEA is closer to the expected behavior of the lattices.
The lattice types \textsc{Octet} and \textsc{BFCC} which have nodes positioned on the body-center and the corners of the unit cells shows a more 'stable' or uniform deformation whereas the \textsc{RhomOcta} and \textsc{TrunOcta} structures tend to buckle more easily.
However, this holds for the prescribed displacement corresponding to a straining of 50\%. In total, the \textsc{TrunOcta} is stiffer and the force necessary for the deformation is higher.

One reason for the deviation of the computed deformation from that of the experiment could be that the finite element model is too stiff. This could have numerical reasons (only a few tetrahedral elements per strut), but it could also be due to the material data. Therefore, we have varied the latter, Fig.~\ref{fig:deformation_abaqus_vf03_2}. For a low modulus, i.e. a very compliant material, also the  \textsc{TrunOcta} structure deforms uniformly. 
In contrast, a non-uniform deformation is observed for the measured modulus and in the same way for a twice as high elastic modulus. Compared with our experimental results, that points to a too stiff finite element model here.


The computed  force-displacement curves are shown in Fig.\,\ref{fig:FEA:force_displacement}. They are of similar shape as the experimental ones but the values of the forces are higher.  The reasons for this deviations are not totally clear to us. In part it may again be attributed to the coarse meshing or the applied contact algorithms which may stiffen the  artificially. Qualitatively, however, we can find a very good agreement between experiment and numerical simulation.


Regarding the stresses in the lattice, all FEA show that the stress peaks are located at the nodes of the structures. This can be explained with the sharp corners here and is in some sense an artifact of the FEA. The overall stress distribution in the deformed specimen is shown in Fig.~\ref{deformation_abaqus_vf02} and Fig.~\ref{deformation_abaqus_vf03} looks very realistically.


\section{Conclusion}\label{sec:conclusion}
%
This paper focuses on a sustainable concept of energy absorption by lattice (micro)structures. The sustainability implies that the structures can withstand multiple loads and return to their original state after a certain recovery time. The different lattice structures are manufactured by SLA printing using an acrylate-based polymeric material, which allows large deformations and behaves tough and viscoelastically. 

We investigated the energy absorption properties of four truss-like lattice types, each with two densities. The lattices differ by their connectivity and topology. 
The increasingly finer resolutions of additive manufacturing technologies make it possible to produce complex components and optimize their internal structure. So, we used structures made of 216 lattice unit cells of 4.5 mm edge length to construct our specimens.
To adapt the lattice to the wanted volume fraction, the thickness of the struts was adjusted so that the investigated lattice structures have a volume fraction of 0.2 and 0.3. In a tension/compression machine, they were then compressed to 50\,\% of their initial height. Their force-displacement diagram was recorded, and the energy absorption was derived. 

The convex \textsc{TrunOcta} lattice, which is based on an octahedron, has proven to be the best. With a mass-specific energy absorption of 1307 J/kg, the \textsc {TrunOcta} structure was able to absorb more than three times as much energy as a standard Octet structure (both with $f_\text{V} = 0.3$).

The \textsc {TrunOcta} has the highest Maxwell number among the lattices investigated, an indicator that its strongly bend-dominated structure is advantageous for energy absorption. Additionally, its thicker struts (of lattices with the same volume fraction) further enhance the  energy absorption capabilities.
Further investigations, e.g., with lattices optimized in that respect, are desirable.

Additionally, we carried out finite element simulations of our experiments. We were able to show that these calculations are, in principle, able to predict the different responses of the lattices. Again, the \textsc{TrunOcta} lattice proved to be the most effective for energy absorption. However, the quantitative evaluation is strongly dependent on the choice of material parameters, making it possible to compare their properties for different lattice structures, but not necessarily to exactly predict their energy absorption.

Finally we emphasize that the specimens here are subjected to a very slow compression and could relax completely. Rapid loading by impact may result in a  different response as we show  in \cite{bieler2024pammsubmitted} and \cite{bieler2024MOMsubmitted}. There we perform experiments with a modified Split-Hopkinson bar setup with the result that most of these specimens were completely destroyed. Thus, all conclusions on the load carrying capacity and on the recovery only apply to a (quasi)static loading regime.

%
%
\backmatter

%
%
%
%
%

\section*{Declarations}

\textbf{Conflict of interest} The authors declare that they have no conflict of interest.

%

%
%
%
%

\begin{appendices}
\section{ }\label{secA1}
%




\begin{table*}[htb]
\caption{Required strut radius $r$ for a given volume fraction $f_\text{V}$ and the resulting cross-section area $A$ for cubic unit cells of $L=4.5$\,mm edge width}
\label{tab:volume_fraction_lattice_structures} 
\begin{tabularx}{\textwidth}{lXXX} \toprule
\noalign{\smallskip}
 & $f_\text{V}$ & $r$ [mm] & $A$ [mm$^2$]\\
\noalign{\smallskip}
\hline
\noalign{\smallskip}
\textsc{Octet} & 0.2 & 0.308 & 145.98\\
\noalign{\smallskip}
\textsc{BFCC} & 0.2 & 0.359 & 122.49\\
\noalign{\smallskip}
\textsc{RhomOcta} & 0.2 & 0.335 & 124.28\\
\noalign{\smallskip}
\textsc{TrunOcta} & 0.2 & 0.442 & 100.46\\
\noalign{\smallskip}
\hline
\noalign{\smallskip}
\textsc{Octet} & 0.3 & 0.390 & 168.65\\
\noalign{\smallskip}
\textsc{BFCC} & 0.3 & 0.456 & 141.36\\
\noalign{\smallskip}
\textsc{RhomOcta} & 0.3 & 0.428 & 142.45\\
\noalign{\smallskip}
\textsc{TrunOcta} & 0.3 & 0.564 & 116.43\\
\noalign{\smallskip}
\hline
\end{tabularx}
\end{table*}

\begin{table*} [htb]
\caption{Number of elements for the FEA of the investigated specimen}
\label{tab:elements} 
\begin{tabularx}{\textwidth}{lXX} \toprule
\noalign{\smallskip}
lattice structure & $f_\text{V}$ & number of elements\\
\noalign{\smallskip}
\hline
\noalign{\smallskip}
\textsc{BFCC} & 0.2 & 282,939\\
\noalign{\smallskip}
\textsc{BFCC} & 0.3 & 410,409\\
\noalign{\smallskip}
\textsc{Octet} & 0.2 & 364,603\\
\noalign{\smallskip}
\textsc{Octet} & 0.3 & 850,594\\
\noalign{\smallskip}
\textsc{RhomOcta} & 0.2 & 364,484\\
\noalign{\smallskip}
\textsc{RhomOcta} & 0.3 & 501,736\\
\noalign{\smallskip}
\textsc{TrunOcta} & 0.2 & 479,854\\
\noalign{\smallskip}
\textsc{TrunOcta} & 0.3 & 657,517\\
\noalign{\smallskip}
\hline
\end{tabularx}
\end{table*}

\end{appendices}



%
\end{document}